\documentclass[twocolumn,floatfix,aps,pre,amsmath]{revtex4}
\usepackage{graphicx}
\usepackage{epsfig}

\begin{document}

\title{Dynamical glass transition: critical temperature $T_c$ and memory kernel in MD-simulated
Ni$_{0.8}$Zr$_{0.2}$}
\author{A.B. Mutiara \thanks{e-mail: amutiara@staff.gunadarma.ac.id}}

\affiliation{Graduate Program in Information System, Gunadarma
University, \\
Jl. Margonda Raya 100, Depok 16424, Indonesian.}

\date{\today}

\begin{abstract}
We use molecular dynamics computer simulations to investigate a
critical temperature $T_c$ for a dynamical glass transition as
proposed by the mode-coupling theory (MCT) of dense liquids in a
glass forming Ni$_{0.8}$Zr$_{0.2}$-system.  The critical
temperature $T_c$ are analyzed from different quantities and
checked the consistency of the estimated values, i.e. from (i) the
non-vanishing nonergodicity parameters as asymptotic solutions of
the MCT equations in the arrested state, (ii) the
${\bf{g}}_m$-parameters describing the approach of the melt
towards the arrested state on the ergodic side, (iii) the
diffusion coefficients in the melt. The resulting $T_c$ values are
found to agree within about 10~$\%$.
\end{abstract}
\maketitle

\section{Introduction}
The transition from a liquid to an amorphous solid that sometimes
occurs upon cooling remains one of the largely unresolved problems
of statistical physics~\cite{goetze04,debenedetti01}. At the
experimental level, the so-called glass transition is generally
associated with a sharp increase in the characteristic relaxation
times of the system, and a concomitant departure of laboratory
measurements from equilibrium. At the theoretical level, it has
been proposed that the transition from a liquid to a glassy state
is triggered by an underlying thermodynamic (equilibrium)
transition~\cite{mezard99}; in that view, an ``ideal'' glass
transition is believed to occur at the so-called Kauzmann
temperature, $T_K$. At $T_K$, it is proposed that only one
minimum-energy basin of attraction is accessible to the system.
One of the first arguments of this type is due to Gibbs and
diMarzio~\cite{gibbs58}, but more recent studies using replica
methods have yielded evidence in support of such a transition in
Lennard-Jones glass formers~\cite{mezard99,coluzzi00a,grigera01}.
These observations have been called into question by experimental
data and recent results of simulations of polydisperse hard-core
disks, which have failed to detect any evidence of a thermodynamic
transition up to extremely high packing fractions~\cite{santen00}.
One of the questions that arises is therefore whether the
discrepancies between the reported simulated behavior of hard-disk
and soft-sphere systems is due to fundamental differences in the
models, or whether they are a consequence of inappropriate
sampling at low temperatures and high densities.

Different, alternative theoretical considerations have attempted
to establish a connection between glass transition phenomena and
the rapid increase in relaxation times that arises in the vicinity
of a theoretical critical temperature (the so-called
``mode-coupling'' temperature, $T_{MCT}$), thereby giving rise to
a ``kinetic'' or ``dynamic'' transition~\cite{goetze92}. In recent
years, both viewpoints have received some support from molecular
simulations. Many of these simulations have been conducted in the
context of models introduced by Stillinger and Weber and by Kob
and Andersen ~\cite{kob95a}; such models have been employed in a
number of studies that have helped shape our current views about
the glass
transition~\cite{coluzzi00a,sastry98,sciortino99,donati99,coluzzi00b,yamamoto00}.

In its simplest (``idealized'') version, firstly analyzed in the ``schematic''
approach by Bengtzelius et al. \cite{bgs} and independently by Leutheusser
\cite{leuth84}, the MCT predicts a transition from a high temperature liquid
(``ergodic'') state to a low temperature arrested (``nonergodic'') state at
a critical temperature $T_c$. Including transversale currents as additional
hydrodynamic variables, the full MCT shows no longer a sharp transition
at $T_c$ but all structural correlations decay in a final $\alpha$-process
\cite{gotzesjo92}. Similar effects are expected from inclusion of
thermally activated matter transport, that means diffusion in the arrested
state \cite{das,sjogren90}.

In the full MCT, the remainders of the transition and the value of $T_c$
have to be evaluated, e.g., from the approach of the undercooled melt towards
the idealized arrested state, either by analyzing the time and temperature
dependence in the $\beta$-regime of the structural fluctuation dynamics
\cite{gleimkob,meyer,cum99} or
by evaluating the temperature dependence of the so-called
${\bf{g}}_m$-parameter \cite{tei96l,tei96e}.
There are further posibilities to estimates $T_c$, e.g., from the temperature
dependence of the diffusion coefficients or the relaxation time of the
final $\alpha$-decay in the melt, as these quantities for $T>T_c$ display
a critical behaviour $|T-T_c|^{\pm \gamma}$. However, only crude estimates
of $T_c$ can be obtained from these quantities, since near $T_c$ the
critical behaviour is masked by the effects of transversale currents and
thermally activated matter transport, as mentioned above.

On the other hand, as emphasized and applied in
\cite{barrat90,mfuchs,kobnauroth}, the value of $T_c$ predicted by
the idealized MCT can be calculated once the partial structure
factors of the system and their temperature dependence are sufficiently
well known. Besides temperature and particle concentration, the partial
structure factors are the only significant quantities which enter the
equations of the so-called nonergodicity parameters of the system.
The latter vanish identically for temperatures above $T_c$
and their calculation thus allows a rather precise determination of the
critical temperature predicted by the idealized theory.

At this stage it is tempting to consider how well the estimates of
$T_c$ from different approaches fit together and whether the $T_c$
estimate from the nonergodicity parameters of the idealized MCT
compares to the values from the full MCT. Regarding this, we here
investigate a molecular dynamics (MD) simulation model adapted to
the glass-forming Ni$_{0.8}$Zr$_{0.2}$ transition metal system.
The Ni$_x$Zr$_{1-x}$-system is well studied by experiments
\cite{kuschke,altounian} and by MD-simulations
\cite{bert1,tei92,teidimat,tei97,teib99,masu1,masu2,masu3,masu4},
as it is a rather interesting system whose components are
important constituents of a number of multi-component 'massive'
metallic glasses. In the present contribution we consider, in
particular, the $x=0.8$ composition and concentrate on the
determination of $T_c$ from evaluating and analyzing the
nonergodicity parameter, the ${\bf{g}}_m(T)$-parameter in the
ergodic regime, and the diffusion coefficients.

Our paper is organized as follows: In section II, we present the
model and give some details of the computations. Section III.
gives a brief discussion of some aspects of the mode coupling
theory as used here. Results of our MD-simulations and their
analysis are then presented and discussed in Section IV.

\section{SIMULATIONS}
The present simulations are carried out as state-of-the-art
isothermal-isobaric ($N,T,p$) calculations. The Newtonian
equations of $N=$ 648 atoms (518 Ni and 130 Zr) are numerically
integrated by a fifth order predictor-corrector algorithm with
time step $\Delta t$ = 2.5 10$^{-15}$s in a cubic volume with
periodic boundary conditions and variable box length L. With
regard to the electron theoretical description of the interatomic
potentials in transition metal alloys by Hausleitner and Hafner
\cite{haushafner}, we model the interatomic couplings as in
\cite{tei92} by a volume dependent electron-gas term $E_{vol}(V)$
and pair potentials $\phi(r)$ adapted to the equilibrium distance,
depth, width, and zero of the Hausleitner-Hafner potentials
\cite{haushafner} for Ni$_{0.8}$Zr$_{0.2}$ \cite{teidimat}. For
this model, simulations were started through heating a starting
configuration up to 2000~K which leads to a homogeneous liquid
state. The system then is cooled continuously to various annealing
temperatures with cooling rate $-\partial_tT$ = 1.5 10$^{12}$~K/s.
Afterwards the obtained configurations at various annealing
temperatures (here 1500-600 K) are relaxed by carrying out
additional isothermal annealing runs. Finally the time evolution
of these relaxed configurations is modelled and analyzed. More
details of the simulations are given in \cite{teidimat}.

\section{THEORY}
\subsection{Nonergodicity parameters}
In this section we provide some basic formulae that permit
calculation of $T_c$ and the nonergodicity parameters $f_{ij}(q)$
for our system. A more detailed presentation may be found in
Refs.~\cite{barrat90,mfuchs,kobnauroth,gotze85,bosse87}. The
central object of the MCT are the partial intermediate scattering
functions which are defined for a binary system by \cite{bernu}
\begin{eqnarray}
F_{ij}(q,t) &=&\frac{1}{\protect\sqrt{N_{i}N_{j}}}\left\langle \rho
^{i}(q,t)\rho ^{j}(-q,0)\right\rangle  \nonumber \\
&=&\frac{1}{\protect\sqrt{N_{i}N_{j}}}\sum\limits_{\alpha
=1}^{N_{i}}\sum\limits_{\beta =i}^{N_{j}} \nonumber \\
&&\times \left\langle \exp (i{\bf{q}}\cdot
[{\bf{r}}_{\alpha }^{i}(t)-{\bf{r}}_{\beta }^{j}(0)])\right\rangle \quad,
\label{T.1}
\end{eqnarray}
where
\begin{equation}
\rho _{i }(\overrightarrow{q})=\sum\limits_{\alpha=1}^{N_{i }}e^{i
\overrightarrow{q}\cdot \overrightarrow{r}_{\alpha i }},\text{ }i =1,2
\label{T.1a}
\end{equation}
is a Fourier component  of the microscopic density of
species $i$.

The diagonal terms $\alpha=\beta$ are denoted as the incoherent intermediate
scattering function
\begin{equation}
F_{i}^{s}(q,t)=\frac{1}{N_{i}}\sum\limits_{\alpha =1}^{N_{i}}\left\langle
\exp (i{\bf{q}}\cdot [{\bf{r}}_{\alpha }^{i}(t)-{\bf{r}}_{\alpha
}^{i}(0)])\right\rangle \quad .
\label{T.2}
\end{equation}

The normalized partial- and incoherent intermediate scattering
functions are given by
\begin{eqnarray}
\Phi_{ij}(q,t)&=& F_{ij}(q,t)/S_{ij}(q) \quad ,\\
\Phi^s_{i}(q,t)&=& F_{i}^{s}(q,t) \quad ,
\end{eqnarray}
where the $S_{ij}(q)= F_{ij}(q,t=0)$ are the partial static structure factors.

The basic equations of the MCT are the set of nonlinear
matrix integrodifferential equations
\begin{equation}
\ddot{{\bf F}}(q,t)+{\mbox{\boldmath $\Omega $}}^2(q){\bf F}(q,t)+
\int_0^td\tau {\bf M}(q,t-\tau) \dot{{\bf F}}(q,\tau) = 0 \quad ,
\label{T.5}
\end{equation}
where ${\bf F}$ is the $2\times 2$ matrix consisting of the
partial intermediate scattering functions  $F_{ij}(q,t)$, and
the frequency matrix ${\mbox{\boldmath $\Omega $}}^2$ is given by
\begin{equation}
\left[{\mbox{\boldmath $\Omega $}}^2(q)\right]_{ij}=q^2k_B T
(x_i/m_i)\sum_{k}\delta_{ik} \left[{\bf S}^{-1}(q)\right]_{kj}\quad.
\label{T.6}
\end{equation}
${\bf S}(q)$ denotes the $2\times 2$ matrix of the
partial structure factors $S_{ij}(q)$, $x_i=N_i/N$ and $m_i$ means the
atomic mass of the species $i$.
The MCT for the idealized glass transition predicts \cite{gotzesjo92}
that the memory kern ${\bf M}$ can be expressed at long times by
\begin{eqnarray}
M_{ij}({\bf q},t)&=&\frac{k_B T}{2\rho m_i x_j}\int\frac{d
{\bf k}}{(2\pi)^3}
\sum_{kl}\sum_{k'l'} \nonumber \\
&& \times V_{ikl}({\bf q},{\bf k}) V_{jk'l'}({\bf q},{\bf q-k}) \nonumber \\
&& \times F_{kk'}({\bf k},t)
F_{ll'}({\bf q-k},t)\quad ,
\label{T.7}
\end{eqnarray}
where $\rho=N/V$ is the particle density and the vertex
$V_{i\alpha\beta}({\bf q},{\bf k})$ is given by
\begin{equation}
V_{ikl}({\bf q},{\bf k})=\frac{{\bf q}\cdot {\bf
k}}{q}\delta_{il} c_{ik}({\bf k})+
\frac{{\bf q}\cdot ({\bf q}-{\bf k})}{q} \delta_{ik} c_{il}
({\bf q}-{\bf k})
\label{T.8}
\end{equation}
and the matrix of the direct correlation function is defined by
\begin{equation}
c_{ij}({\bf q})=\frac{\delta_{ij}}{x_i}-
\left[{\bf S}^{-1}({\bf q})\right]_{ij} \quad .
\label{T.9}
\end{equation}

The equation of motion for $F^s_i(q,t)$ has a similar form as
Eq.(\ref{T.5}), but the memory function for the incoherent
intermediate scattering
function is given by
\begin{eqnarray}
M_{i}^{s}({\bf q},t) & = & \int \frac{d{\bf k}}{(2\pi)^3} \frac{1}{\rho}
\left(\frac{{\bf q}\cdot {\bf k}}{q}\right) (cF)_i ({\bf k},t)
\nonumber \\
&& \times F_{i}^{s}({\bf q}-{\bf k},t) ,
\label{T.10}
\end{eqnarray}
\begin{eqnarray}
(cF)_i(k,t)&=&(c_{ii}(q))^2 F_{ii}(q,t)+2c_{ii}(q)c_{ij}(q)F_{ij}(q,t)
\nonumber \\
&& +(c_{ij}(q))^2F_{jj}(q,t)\quad j\neq i \quad .
\label{T.11}
\end{eqnarray}

In order to characterize the long time behaviour of
the intermediate scattering function, the nonergodicity parameters
${\bf f}({\bf q})$ are introduced as
\begin{equation}
f_{ij}({\bf q})=lim_{t\to \infty}\Phi_{ij}({\bf q},t) \quad .
\label{T.12}
\end{equation}
These parameters are the solution of
eqs.~(\ref{T.5})-(\ref{T.9}) at long times. The meaning of these parameters
is the following:
if $f_{ij}({\bf q})=0$, then the system is
in a liquid state with
density fluctuation correlations decaying at long times. If
$f_{ij}({\bf q})>0$, the system is in an  arrested, nonergodic state, where
density fluctuation correlations are stable for all times.
In order to compute $f_{ij}({\bf q})$, one can use the following iterative
procedure~\cite{kobnauroth}:
\begin{eqnarray}
{\bf f}^{(l+1)}(q) &=& \frac{
{\bf S}(q) \cdot {\bf N}[{\bf f}^{(l)},{\bf f}^{(l)}](q) \cdot {\bf S}
(q)}{{\bf Z}} \nonumber \\
&& + \frac{q^{-2}|{\bf S}(q)| |{\bf N}[{\bf f}^{(l)},{\bf f}^{(l)}](q)|
{\bf S}(q)}{\bf Z} \quad ,
\label{T.13}
\end{eqnarray}
\begin{eqnarray}
{\bf Z}&=&
q^2+Tr({\bf S}(q) \cdot {\bf N}[{\bf f}^{(l)},{\bf f}^{(l)}](q)) \nonumber \\
&& + q^{-2}| {\bf S}(q)| | {\bf N}[{\bf f}^{(l)},{\bf f}^{(l)}](q)|
\nonumber \quad,
\end{eqnarray}
where the matrix ${\bf N}(q)$ is given by
\begin{equation}
N_{ij}(q)=\frac{m_i}{x_i k_B T} M_{ij}(q) \quad.
\label{T.14}
\end{equation}

This iterative procedure, indeed,  has two type of solutions,
nontrivial ones with ${\bf f}(q)>0$ and trivial solutions ${\bf f}(q)=0$.

The incoherent nonergodicity parameter $f_i^{s}(q)$ can be evaluated
by the following iterative procedure:

\begin{equation}
q^2 \frac{f_i^{s,l+1}(q)}{1-f_i^{s,l+1}(q)} = M_i^{s}[{\bf f},
f_i^{s,l}](q)
\quad .
\label{T.15}
\end{equation}

As indicated by Eq.(\ref{T.15}), computation of the incoherent
nonergodicity parameter $f_i^s(q)$ demands that
the coherent nonergodicity parameters are determined in advance.

\subsection{$\bf{g}_m$--parameter}

Beyond the details of the MCT, equations of motion like (\ref{T.5})
can be derived
for the correlation functions under rather
general assumptions within the Lanczos recursion scheme \cite{lanczos}
resp. the Mori-Zwanzig formalism \cite{morizwanzig}.
The approach demands that the time dependence of fluctuations
A, B, ... is governed by a time evolution operator like the
Liouvillian and that for two fluctuating quantitites a scalar
products (B, A) with the meaning of a correlation function can
be defined. In case of a tagged particle, this leads
for $\Phi^s_i(q,t)$ to the exact equation
\begin{equation}
\ddot{\Phi}^s_i(q,t)/\Omega_{0}^{2}+\Phi^s_i(q,t)
+\int_{0}^{t}d\tau
M^0_i(q,t-\tau )\dot{\Phi}^s_i(q,\tau )=0
\label{G.1}
\end{equation}
\noindent with memory kernel $M^0_i(q,t)$ in terms of
a continued fraction.

Within $M^0_i(q,t)$ are hidden all the details of
the time evolution of $\Phi^s_i(q,t)$. As proposed and
applied in \cite{tei96l,tei96e}, instead of calculating
$M^0_i(q,t)$ from the time evolution operator as a
continued fraction, it can be evaluated in closed forms
once $\Phi^s_i(q,t)$ is known, e.g., from experiments
or MD-simulations.
This can be demonstrated by introduction of
\begin{eqnarray}
\Phi _{c}(\omega )\pm i\Phi _{s}(\omega )&:=&
\lim_{\varepsilon \rightarrow 0}{\mathcal{L}}{\left\{ \Phi\right\}}
(\varepsilon\mp i\omega) \quad,
\label{G.2}
\end{eqnarray}
\noindent with
\begin{equation}
{\mathcal{L}}{\left\{ \Phi\right\}}(z)=
\int_{0}^{\infty}dt e^{-zt}\Phi(t)
\label{G.3}
\end{equation}
\noindent the Laplace transform of $\Phi(t)$, and
\begin{eqnarray}
M^0_i(\omega)_c \pm i M^0_i(\omega)_s &:=& \lim_{\varepsilon
\rightarrow 0}{\mathcal{L}}{\left\{ M^0_i\right\}} (\varepsilon\mp
i\omega) \quad. \label{G.4}
\end{eqnarray}
Eq.(\ref{G.1}) then leads to
\begin{equation}
M^0_i(\omega)_c=\frac{\Phi_{c}(\omega)}{\left[ 1-\omega
\Phi_{s}(\omega )\right] ^{2}+\left[ \omega
\Phi _{c}(\omega )\right] ^{2}} \quad.
\label{G.5}
\end{equation}
On the time axis, $M^0_i(t)$ is given by
\begin{equation}
M^0_i(t)=\frac{2}{\pi }\int_{0}^{\infty }d\omega M^0_i(\omega)_c
\cos (\omega t) \quad.
\label{G.6}
\end{equation}

Adopting some arguments from the schematic MCT, Eq.(\ref{G.1}) allows
asymptotically finite correlations $\Phi^s_i(q,t\rightarrow\infty)>0$,
that means an arrested state, if $M^0_i(q,t\rightarrow\infty)$ remains finite where
the relationship holds
\begin{equation}
M^0_i(q,t\rightarrow\infty)
(\Phi^s_i(q,t\rightarrow\infty)^{-1}- 1)=1 \quad.
\label{G.7}
\end{equation}
In order to characterize the undercooled melt and its transition into
the glassy state, we introduced in \cite{tei96l} the function
\begin{equation}
{\bf{G}}(\Phi,M^0):= M^0(t)(1/\Phi(t)- 1) \quad.
\label{G.8}
\end{equation}

According to (\ref{G.7}), ${\bf{G}}(\Phi,M^0)$ has the property that
\begin{equation}
{\bf {G}}(\Phi,M^0)\mid_{t\rightarrow\infty}= 1
\label{G.9}
\end{equation}
\noindent in the arrested, nonergodic state. On the other hand,
if
\begin{equation}
{\bf {g}}_m:=Max\left \{{\bf {G}}(\Phi,M^0)\mid 0<t<\infty \right\} < 1\quad,
\label{G.10}
\end{equation}
there is no arrested solution and the correlations
$\Phi^s_i(q,t)$ decay to zero for $t\rightarrow\infty$, that
means, the system is in the liquid state. From that we proposed
\cite{tei96l} to use the value of ${\bf{g}}_m$ as a relative
measure how much the system has approached the arrested state and
to use the temperature dependence of ${\bf{g}}_m(T)$ in the
liquid state as an indication how the system approaches this state.

\section{Results and Discussions \label{RD}}

\subsection{Partial structure factors and intermediate scattering functions}
First we show the results of our simulations concerning
the static properties of the system in terms of the partial structure factors
$S_{ij}(q)$ and partial correlation functions $g_{ij}(r)$.

To compute the partial structure factors $S_{ij}(q)$
for a binary system we use the following
definition \cite{hansen}
\begin{eqnarray}
S_{ij }(\overrightarrow{q}) &=&x_{i}\delta _{ij
}+\rho x_{i}x_{j}\int (g_{ij}(r)-1)e^{-i
\overrightarrow{q}\cdot \overrightarrow{r}}d\overrightarrow{r}
\label{E.5} \quad,
\end{eqnarray}
where
\begin{equation}
g_{ij }(\overrightarrow{r})=\frac{V}{N_{i}N_{j}}
\left\langle \sum\limits_{\alpha=1}^{N_{i}}\sum_{\beta=1,\beta\neq \alpha}^
{N_{j}}\delta ({\bf{r}}-\left| {\bf{r}}_{\alpha}(t)-
{\bf{r}}_{\beta}(t)\right| )\right\rangle
\label{E.5a}
\end{equation}
are the partial pair correlation functions.

The MD simulations yield a periodic repetition of the atomic distributions
with periodicity length $L$. Truncation of the
Fourier integral in Eq.(\ref{E.5}) leads to an
oscillatory behavior of
the partial structure factors at small $q$. In order to reduce
the effects  of this truncation, we compute from Eq.(\ref{E.5a})
the partial pair correlation functions for distance $r$ up to $R_c=3/2L$.
For numerical evaluation of Eq.(\ref{E.5}), a Gaussian type damping term
is included
\begin{eqnarray}
S_{i j }(q)&=&x_{i }\delta _{i j }+4\pi \rho x_{i
}x_{j }\int\limits_{0}^{R_{c}}r^{2}(g_{i j }(r)-1)\frac{\sin
(qr)}{qr} \nonumber \\
&& \times \exp (-(r/R)^{2})dr
\label{E.8GS}
\end{eqnarray}
with $R=R_{c}/3$.

\begin{figure}[htbp]
\centering
\psfig{file=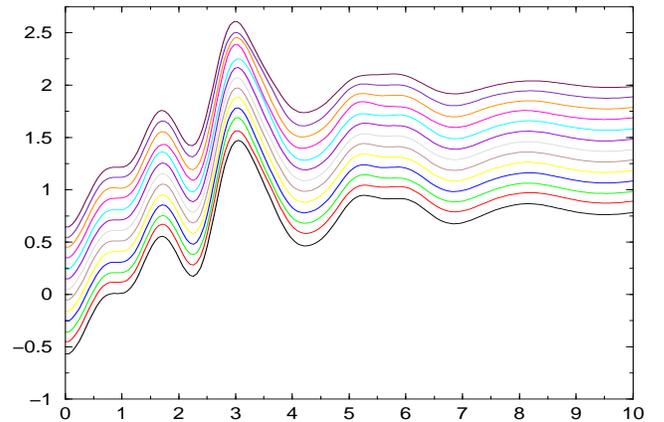,width=8.5cm,height=5.50cm}
\caption{Partial structure factors Ni-Ni-part at $T=$ 1500~K,
1400~K, 1300~K, 1200~K, 1100~K, 1000~K, 900~K and 800~K (from top
to bottom),the curves are vertically shifted by 0.05 relative to
each other. } \label{fig1}
\end{figure}

Fig.\ref{fig1}- fig.\ref{fig2a} shows the partial structure
factors $S_{ij}(q)$ versus $q$ for all temperatures investigated.
The figure indicates that the shape of $S_{ij}(q)$ depends weakly
on temperature only and that, in particular, the positions of the
first maximum and the first minimum in $S_{ij}(q)$ are more or
less temperature independent.

\begin{figure}[htbp]
\centering
\psfig{file=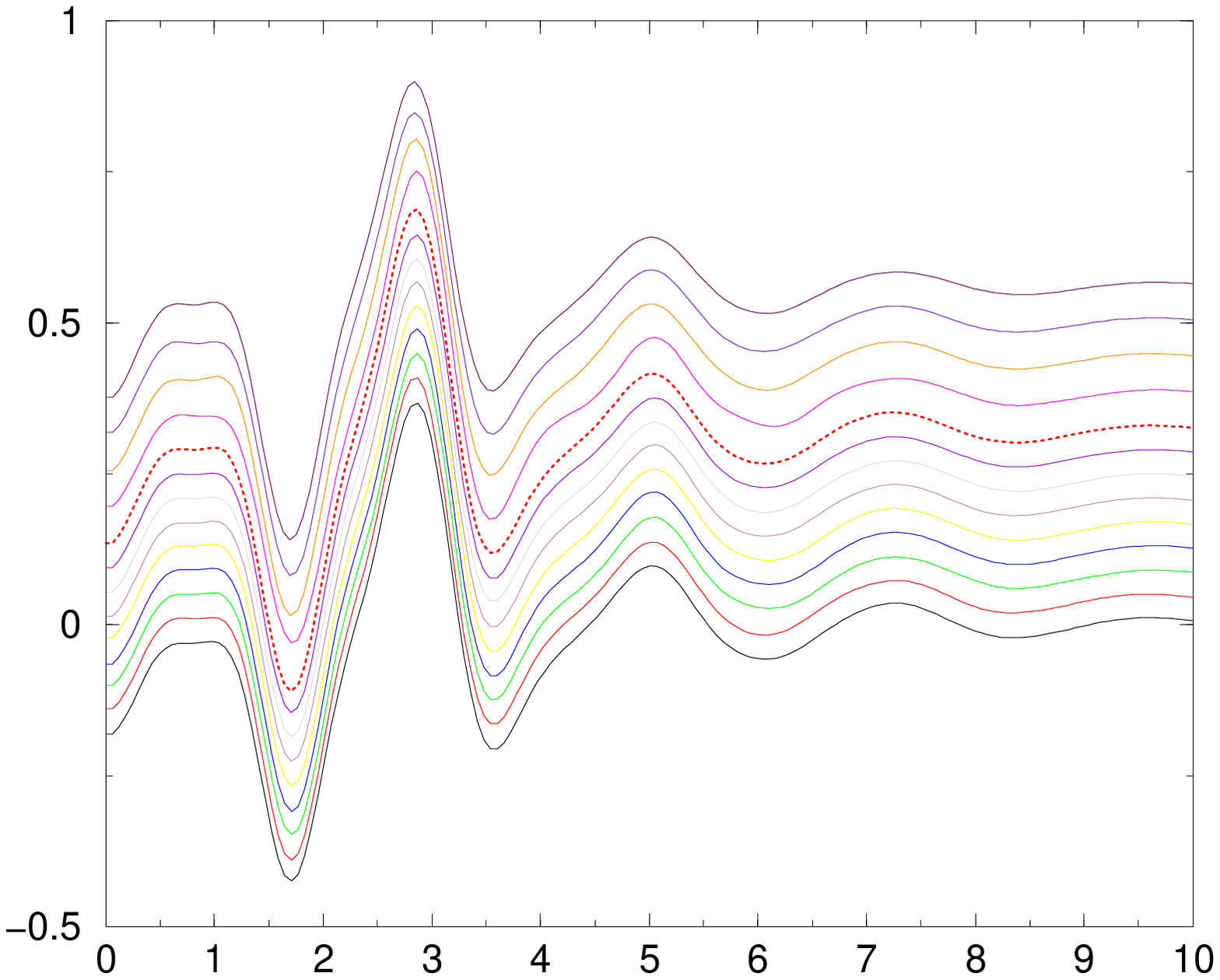,width=8.5cm,height=5.50cm}
\caption{Partial structure factors Ni-Zr-part at $T=$ 1500~K, $T=$
1400~K, 1300~K, 1200~K, 1100~K, 1000~K, 900~K and 800~K (from top
to bottom),the curves are vertically shifted by 0.05 relative to
each other. } \label{fig1a}
\end{figure}

\begin{figure}[htbp]
\psfig{file=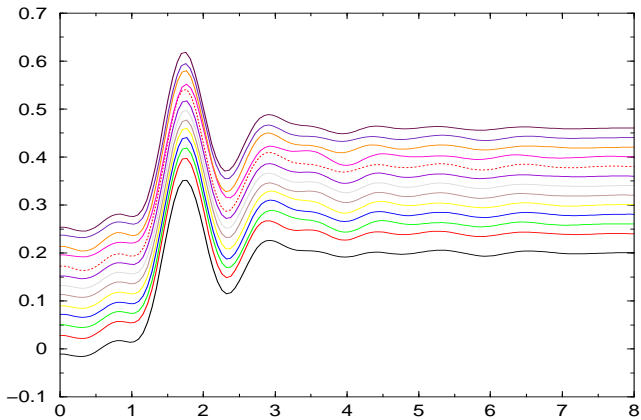 ,width=8.5cm,height=5.50cm}
\caption{Partial structure factors Zr-Zr-part at $T=$ 1500~K, $T=$
1400~K, 1300~K, 1200~K, 1100~K, 1000~K, 900~K and 800~K (from top
to bottom),the curves are vertically shifted by 0.05 relative to
each other.} \label{fig2a}
\end{figure}

To investigate the dynamical properties of the system, we have
calculated the incoherent scattering function $F^s_{i}(q,t)$ and
the coherent scattering function $F_{ij}(q,t)$ as defined in
equations (\ref{T.1}) and (\ref{T.2}).

\begin{figure}[htbp]
\psfig{file=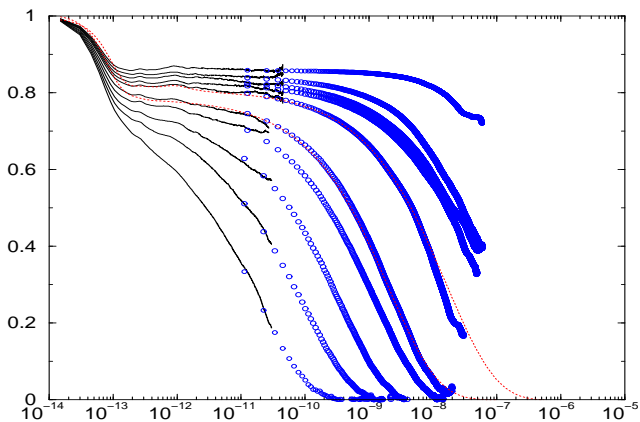,width=8.5cm,height=5.50cm}
\caption{Incoherent intermediate scattering function
$\Phi^s_i(q,t)$ Ni-part for $q=24.4$~nm$^{-1}$ at $T=$ 1500~K,
1400~K, 1300~K, 1200~K, 1100~K, 1000~K, 950~K, 900~K, and 800~K
(from left to right).} \label{fig2b}
\end{figure}

\begin{figure}[htbp]
\psfig{file=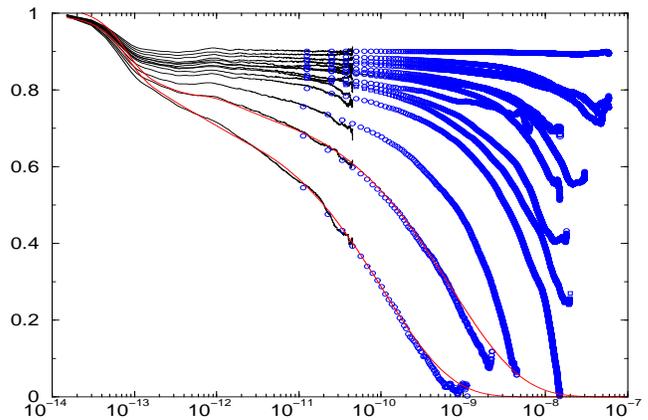,width=8.5cm,height=5.50cm}
\caption{The same as fig.\ref{fig2b} but for Zr-part.}
\label{fig3a}
\end{figure}

Fig.\ref{fig2b} and fig.\ref{fig3a} presents the normalized
incoherent intermediate scattering functions $\Phi^s_i(q,t)$ of
both species evaluated from our MD data for wave vector
$q_n$=$2\pi n/L$ with n = 9, that means $q_9=24.4$ nm~$^{-1}$.
From the figure we see that $\Phi^s_i(q,t)$ of both species shows
at intermediate temperatures a structural relaxation in three
succesive steps as predicted by the idealized schematic MCT
\cite{hansenyip}. The first step is a fast initial decay on the
time scale of the vibrations of atoms ($t<0.2$~ps). This step is
characterized by the MCT only globaly. The second step is the
$\beta $-relaxation regime. In the early $\beta$-regime the
correlator should decrease according to
$\Phi^s_i(q,t)=f_{csi}(q)+A/t^{a}$ and in the late
$\beta$-relaxation regime, which appears only in the melt,
according the von Schweidler law $f_{csi}(q)-Bt^{b}.$ Between them
a wide plateau is found near the critical temperature $T_{c}$. In
the melt, the $\alpha$-relaxation takes place as the last decay
step after the von Schweidler-law. It can be described by the
Kohlrausch-Williams-Watts (KWW) law $\Phi^s_i(q,t)=A_{0}\exp
(-(t/\tau _{\alpha})^{\beta })$ where the relaxation time $\tau
_{\alpha}$ near the glass transition shifts drastically to longer
times.

The inverse power-law decay for the early
$\beta$-regime $\Phi \sim f_{c}+A/t^{a}$ is not seen
in our data. This seems to be due to the fact
that in our system the power-law decay is dressed by the atomic vibrations
(\cite{tei96l,tei96e} and references therein).

According to our MD-results, $\Phi^s_i(q,t)$ decays
to zero for longer times at all temperatures
investigated. This is in agreement with the full MCT.
Including transversal currents as additional hydrodynamic
variables, the full MCT \cite{gotzesjo92} comes to
the conclusion that all structural correlations
decay in the final $\alpha$-process, independent of
temperature. Similar effects are expected from inclusion of
thermally activated matter transport, that means diffusion in the arrested
state.

At $T=$~900~K - 700~K, the $\Phi^s_i(q,t)$ drop rather sharply at
large $t$. This reflects aging effects which take place, if a
system is in a transient, non-steady state \cite{kobbarrat}. Such
a behaviour indicates relaxations of the system on the time scale
of the 'measuring time' of the correlations.

\subsection{Nonergodicity parameters}

\begin{figure}[htbp]
\psfig{file=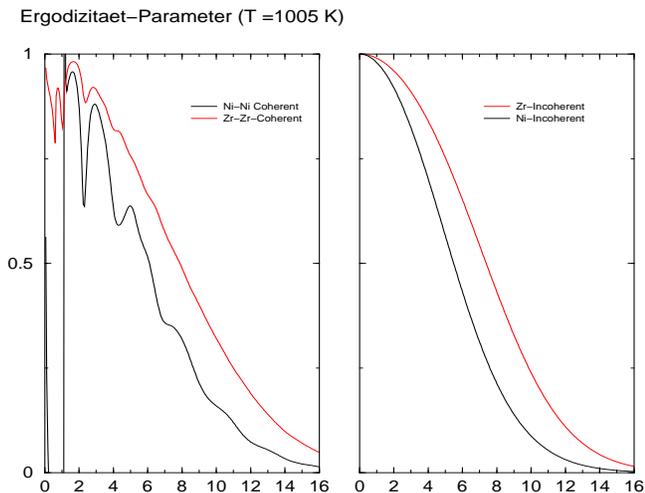,width=8.5cm,height=6.50cm}
\caption{Non-ergodicity parameter $f_{cij}$ for the incoheremt and
coherent intermediate scattering functions as solutions of eqs.
(\ref{T.6}) and (\ref{T.7})}
\label{fig3b}
\end{figure}

The nonergodicity parameters are defined by Eq.(\ref{T.12}) as a
non-vanishing asymptotic solution of the MCT Eq.(\ref{T.5}).
Fig.~\ref{fig3b} presents the estimated $q$-dependent
nonergodicity parameters from the coherent and incoherent
scattering functions of Ni and Zr at T=1005~K.

In order to compute the nonergodicity parameters $f_{ij}(q)$
analytically, we followed for our binary system the self-consistent
method as formulated by Nauroth and Kob \cite{kobnauroth} and
as sketched in Section III.A. Input data for our
iterative determination of $f_{ij}(q) = F_{ij}(q,\infty)$ are
the temperature dependent partial structure factors $S_{ij}(q)$
from the previous subsection. The iteration is started
by arbitrarily setting $F_{Ni-Ni}(q,\infty)^{(0)}=0.5 S_{Ni-Ni}(q)$,
$F_{Zr-Zr}(q,\infty)^{(0)}=0.5 S_{Zr-Zr}(q)$,
$F_{Ni-Zr}(q,\infty)^{(0)}=0$.

For $T > 1100$~K we always obtain the trivial solution $f_{ij}(q)
= 0$ while at T = 1000 K and below we get stable non-vanishing
$f_{ij}(q)>0$. The stability of the non-vanishing solutions was
tested for more than 3000 iteration steps. From this results we
expect that $T_c$ for our system lies between 1000 and 1100 K. To
estimate $T_c$ more precisely, we interpolated $S_{ij}(q)$ from
our MD data for temperatures between 1000 and 1100 K by use of the
algorithm of Press et.al. \cite{press}. We observe that at $T =
1005$~K a non-trivial solution of $f_{ij}(q)$ can be found, but
not at $T = 1010$~K and above. It means that the critical
temperature $T_c$ for our system is around 1005 K. The non-trivial
solutions $f_{ij}(q)$ for this temperature shall be denoted the
critical nonergodicty parameters $f_{cij}(q)$. They are included
in Fig.~\ref{fig3b}.

By use of the critical nonergodicity parameters $f_{cij}(q)$, the
computational procedure was run to determine the critical
nonergodicity parameters $f^s_{ci}(q)$ for the incoherent
scattering functions at T = 1005 K . Fig.~\ref{fig3b} also
presents our results for the so calculated $f^s_{ci}(q)$.

\subsection{${\bf{g}}(\Phi^s_i,M^0_i)$-function and ${\bf{g}}_m$-parameters}

Here we present our results about the
${\bf{g}}(\Phi^s_i,M^0_i)$-function \cite{tei96l,tei96e} described
in section III.B. The memory functions $M_i^0(q,t)$  are evaluated
from the MD data for $\Phi_i^s(q,t)$ by Fourier transformation
along the positive time axis. For completeness, also $T=700$ and
800 K data are included where the corresponding  $\Phi_i^s(q,t)$
are extrapolated to longer times by use of an KWW approximation.

\begin{figure}[htbp]
\psfig{file=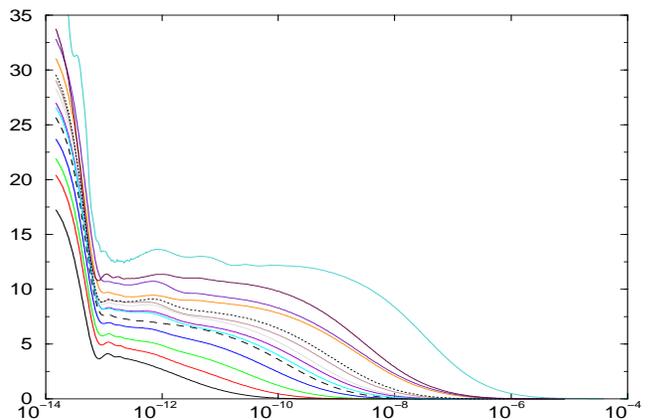,width=8.5cm,height=5.50cm}
\caption{Ni-Part: Time dependence of the dimensionless memory
function $M^0_s(q,t)/\Omega^2_{s-i}$ from MD simulations for
$q_9=21.6$~nm$^{-1}$ and $T=$~800~K, 900~K, 950~K, 1000~K, 1100~K,
1200~K, 1300~K, 1400~K, and 1500~K (from top to bottom) }
\label{fig4a}
\end{figure}

\begin{figure}[htbp]
\psfig{file=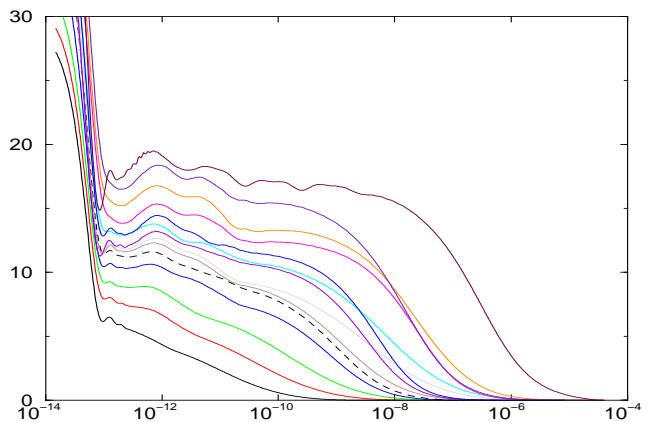,width=8.5cm,height=5.50cm}
\caption{The same as fig.\ref{fig4a} but for Zr-part. }
\label{fig4b}
\end{figure}

Fig.~\ref{fig4a} and Fig.~\ref{fig4b} show the thus deduced
$M_i^0(q,t)$ for $q = 24.4$~nm$^{-1}$. Regarding their qualitative
features, the obtained  $M_i^0(q,t)$ are in full agreement with
the results in \cite{tei96e} for the Ni$_{0.5}$Zr$_{0.5}$ system.
A particular interesting detail is the fact that there exists a
minimum in $M_i^0(q,t)$ for both species, Ni and Zr, at all
investigated temperatures around a time of 0.1 ps. Below this
time, $\Phi_i^s(q,t)$ reflects the vibrational dynamics of the
atoms. Above this value, the escape from the local cages takes
place in the melt and the $\beta$-regime dynamics are developed.
Apparently, the minimum is related to this crossover.

\begin{figure}[htbp]
\psfig{file=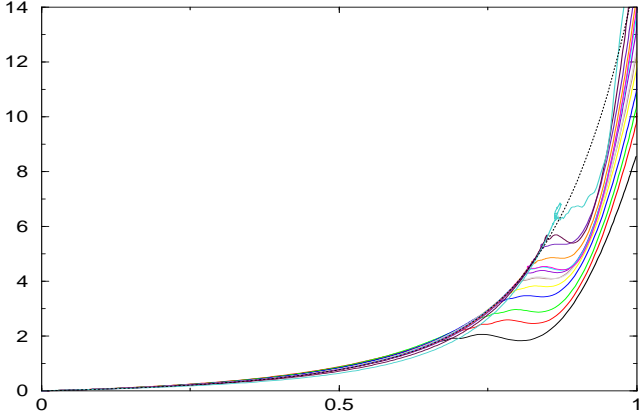,width=8.5cm,height=5.50cm}
\caption{Ni-Part: Dimensionless memory function
$M^0_s(q,t)/\Omega^2_{s-i}$ as a function of $\Phi^s_i(q,t)$
(solid line) for $q_9=24.4$~nm$^{-1}$ and $T=$~800~K, 900~K,
950~K, 1000~K, 1100~K, 1200~K, 1300~K, 1400~K, and 1500~K (from
top to bottom); a) Ni-part and b) Zr-part ; Polynom fit of the low
$\Phi$ memory function $M(\Phi)$ at $T=800$~K (long dashed line).}
\label{fig5}
\end{figure}

\begin{figure}[htbp]
\psfig{file=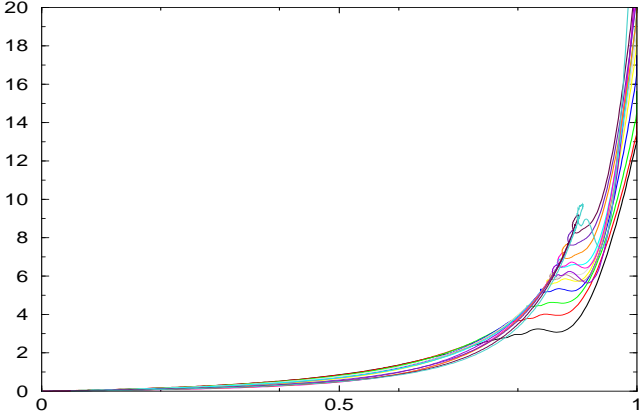,width=8.5cm,height=5.50cm}
\caption{The same as fig.\ref{fig5} but for Zr-part.}
\label{fig5a}
\end{figure}

In Fig.~\ref{fig5} and Fig.~\ref{fig5a} we display $M_i^0(q,
\Phi_i^s(q,t))$, that means $M_i^0(q,t)$ versus  $\Phi_i^s(q,t)$.
In this figure we again find the features already described for
Ni$_{0.5}$Zr$_{0.5}$ in \cite{tei96l,tei96e}. According to the
plot, there exist ($q$-dependent) limiting values
 $\Phi_{i0}^s(q,t)$ so that $M_i^0(q,t)$ for
$\Phi_i^s(q,t)< \Phi_{i0}^s(q,t)$ is close to an universal
behavior, while for $\Phi_i^s(q,t)> \Phi_{i0}^s(q,t)$ marked
deviations are seen. $\Phi_{i0}^s(q,t)$ significantly decreases
with increasing temperature. It is tempting to identify
$M_i^0(q,t)$  below  $\Phi_{i0}^s(q,t)$ with the polynomial form
for  $M_i^0(q,t)$ assumed in the schematic version of the MCT
\cite{gotzesjo92}. In fig.~\ref{fig5} and fig.~\ref{fig5a}, the
polynomial obtained by fitting the 1000 K data below
$\Phi_{i0}^s(q,t)$ is included by a dashed line, extrapolating it
over the whole $\Phi$-range.

\begin{figure}[htbp]
\psfig{file=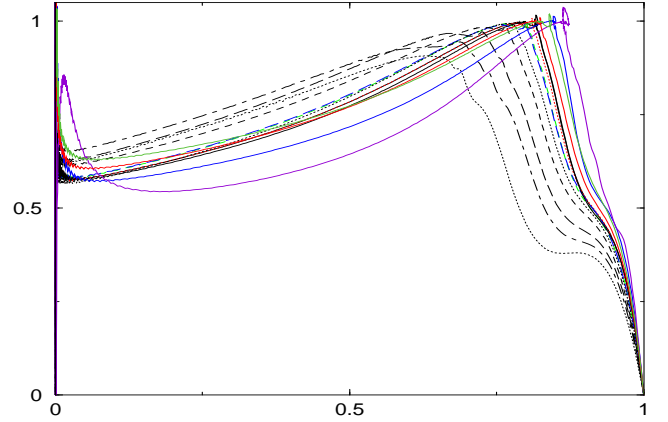,width=8.5cm,height=5.50cm}
\caption{Ni-part: MD simulation results for the characteristic
function ${\bf{g}}(\Phi^s)$ as a function of $\Phi^s_i$ for
$q=24.4$~nm$^{-1}$} \label{fig6}
\end{figure}

\begin{figure}[htbp]
\psfig{file=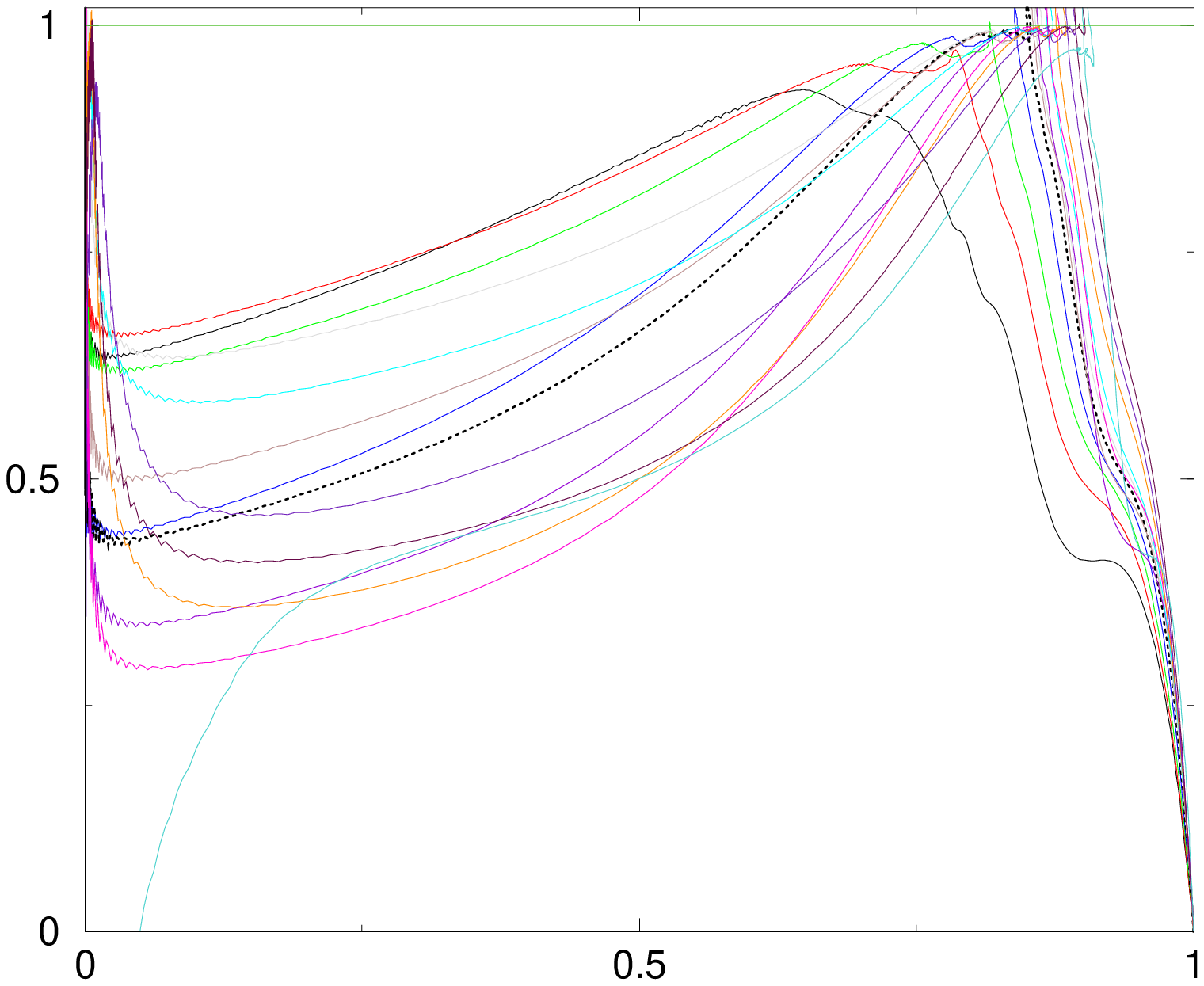,width=8.5cm,height=5.50cm}
\caption{The same as fig.\ref{fig6} but for Zr-part.} \label{fig7}
\end{figure}

\begin{figure}[htbp]
\psfig{file=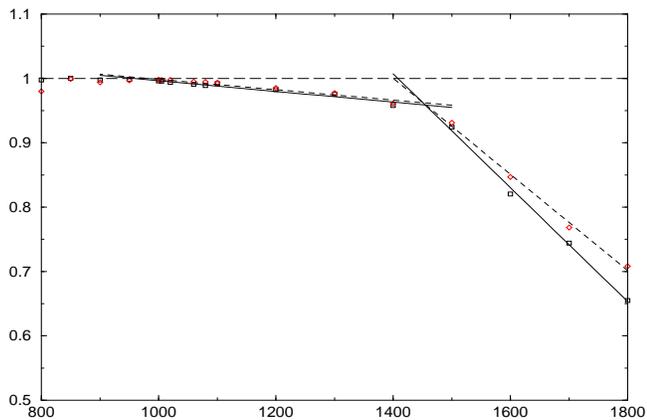,width=5.50cm,height=8.5cm,angle=270}
\caption{MD simulation results of the temperature dependence of
${\bf{g}}_m(q,T)$ for $q_9=24.4$~nm$^{-1}$ (symbols). Linear fits
to the ${\bf{g}}_m(q,T)$ are included by full and dash lines ( for
$q_9=24.4$~nm$^{-1}$); a) Zr-part with $T_c=970$~K and b) Ni-part
with $T_c=950$~K. } \label{fig8}
\end{figure}

By use of the calculated memory functions, we can evaluate the
${\bf{g}}(\Phi^s_i,M^0_i)$, Eq.(\ref{G.8}). In Fig.\ref{fig6} and
Fig.~\ref{fig7} this quantity is presented versus the
corresponding value of $\Phi_i^s(q,t)$ and denoted as
${\bf{g}}(\Phi^s_i)$. For all the investigated temperatures,
${\bf{g}}(\Phi^s_i)$ has a maximum ${\bf{g}}_m(q,T)$ at an
intermediate value of $\Phi$. In the high temperature regime, the
values of ${\bf{g}}_m(q,T)$ move with decreasing temperature
towards the limiting value 1. This is, in particular, visible in
Fig.~\ref{fig8} where we present ${\bf{g}}_m(q,T)$ as function of
temperature for both species, Ni and Zr, and wave-vectors
$q_9=24.4$~nm$^{-1}$. At temperatures above 1000 K, the
${\bf{g}}_m$-values increase approximately linear towards 1 with
decreasing temperatures. Below 1000 K, they remain close below the
limiting value of 1, a behavior denoted in \cite{tei96l,tei96e} as
a balancing on the borderline between the arrested and the
non-arrested state due to thermally induced matter transport by
diffusion in the arrested state at the present high temperatures.

Linear fit of the ${\bf{g}}_m$-values for Ni above 950~K and for
Zr above 1000~K predicts a crossover temperature $T^*_c$ from
liquid (${\bf{g}}_m < 1$) to the quasi-arrested (${\bf{g}}_m = 1$)
behavior around 970 K from the Ni data and around 1020 K from the
Zr data. We here identify this crossover temperature with the
value of $T_c$ as visible in the ergodic, liquid regime and
estimate it by the mean value from the Ni- and Zr-subsystems, that
means by $T_c = 1000$~K.

While in \cite{tei96l,tei96e} for the Ni$_{0.5}$Zr$_{0.5}$ melt a
$T_c$-value of 1120~K was estimated from ${\bf{g}}_m(T)$, the value
for the present composition is lower by about 120~K. A significant
composition dependence of $T_c$ is expected according to the results
of MD simulation for the closely related Co$_x$Zr$_{1-x}$ system
\cite{teirossler}. Over the whole $x$-range, $T_c$ was found to vary
between 1170 and 650~K in Co$_x$Zr$_{1-x}$, with $T_c$($x=0.2$) $\simeq$
800~K. Regarding this, the present data for the Ni$_x$Zr$_{1-x}$ system
reflect a rather weak $T_c$ variation.

\subsection{Diffusion-coefficients}

From the simulated atomic motions in the computer experiments, the
diffusion coefficients of the Ni and Zr species can be determined
as the slope of the atomic mean square displacements in the
asymptotic long-time limit
\begin{equation}
D_{i}(T)=\lim\limits_{t\rightarrow \infty }\frac{(1/N_{i})\sum\limits_{%
\alpha =1}^{N_{i}}\left| \mathbf{r}_{\alpha }(t)-\mathbf{r}_{\alpha
}(0)\right| ^{2}}{6t} \quad.
\end{equation}

\begin{figure}[htbp]
\psfig{file=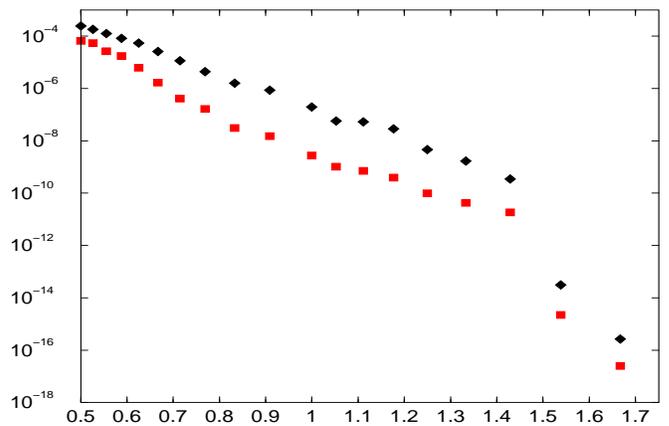,width=8.5cm,height=5.50cm}
\caption{Diffusion coefficients $D_i$ as a function of $1000/T$.
Symbols are MD results for Zr (squares) and Ni (diamonds)}
\label{fig9}
\end{figure}

Fig.~\ref{fig9} shows the thus calculated diffusion coefficients
of our Ni$_{0.8}$Zr$_{0.2}$ model for the temperature range
between 600 and 2000 K. At temperatures above approximately 1000
K, the diffusion coefficients for both species run parallel to
each other in the Arrhenius plot, indicating a fixed ratio
$D_{Ni}/D_{Zr}\approx  2.5$ in this temperature regime. At lower
temperatures, the Ni atoms have a lower mobility than the Zr
atoms, yielding around 800 K a value of about 10 for
$D_{Ni}/D_{Zr}$. That means, here the Zr atoms carry out a rather
rapid motion within a relative immobile Ni matrix.

According to the MCT, above $T_c$ the diffusion coefficients follow
a critical power law
\begin{equation}
D_{i}(T)\sim (T-T_{c})^{\gamma }, \text{ for }T > T_c \label{MA.1}
\end{equation}
with non-universal exponent $\gamma$ \cite{kob951}.

In order to estimate $T_c$ from this relationship, we have adapted
the critical power law by a least mean squares fit to the
simulated diffusion data for 1000 K and above. According to this
fit, the system has a critical temperature of about 850-900 K.


Similar results for the temperature dependence of the diffusion coefficients
have been found in MD simulations for other metallic glass forming systems,
e.g., for Ni$_{0.5}$Zr$_{0.5}$ \cite{teidimat}, for Ni$_{x}$Zr$_{1-x}$
\cite{teirossler}, Cu$_{0.33}$Zr$_{0.67}$ \cite{gaukel}, or
Ni$_{0.81}$B$_{0.19}$ \cite{vane}. In all cases, like here, a break
is observed in the Arrhenius slope. In the mentioned Zr-systems, this break
is related to a change of the atomic dynamics around $T_c$ whereas for
Ni$_{0.81}$B$_{0.19}$ system it is ascribed to $T_G$.
As in \cite{vane} $T_c$ and $T_G$ apparently fall together, there is
no serious conflict between the obervations.

\section{Conclusion}
The present contribution reports results from MD simulations of a
Ni$_{0.8}$Zr$_{0.2}$ computer model. The model is based on the
electron theoretical description of the interatomic potentials for
transition metal alloys by Hausleitner and Hafner
\cite{haushafner}. There are no parameters in the model adapted to
the experiments.

There is close agreement between the $T_c$ values estimated from
the dynamics in the undercooled melt when approaching $T_c$ from
the high temperature side. The values are $T_c \approx 950 -
1020$~K from the ${\bf{g}}_m$-parameters, and $T_c \approx 950$~K
from the diffusion coefficients. As discussed in
\cite{teirossler}, the $T_c$-estimates from the diffusion
coefficients seem to depend on the upper limit of the temperature
region taken into account in the fit procedure, where an increase
in the upper limit increases the estimated $T_c$. Accordingly,
there is evidence that the present value of 950 K may
underestimate the true $T_c$ by about 10 to 50 K, as it based on
an upper limit of 2000 K only. Taking this into account, the
present estimates from the melt seem to lead to a $T_c$ value
around 1000 K.

The $T_c$ from the nonergodicity parameters describe the approach
of the system towards $T_c$ from the low temperature side. They
predict a $T_c$ value of 1005 K. This value is clearly outside the
range of our $T_c$ estimates from the high temperature, ergodic
melt. We consider this as a significant deviation which, however,
is much smaller than the factor of two found in the modelling of a
Lennard-Jones system \cite{kobnauroth}. The here observed
deviation between the $T_c$ estimates from the ergodic and the
so-called nonergodic side reconfirm the finding from the soft
spheres model\cite{mfuchs} of an agreement within some 10 $\%$
between the different $T_c$-estimates.

\begin{acknowledgments}
A.B.M. gratefully  acknowledges financial support
of the SFB 602 during the post-doctoral program.
\end{acknowledgments}

\end{document}